\documentclass[a4paper,fleqn,usenatbib]{mnras}

\newcommand{\Msun}{M$_{\odot}$ }

\newcommand{\Mjup}{M$_{\mathrm{JUP}}$ }

\usepackage[T1]{fontenc}
\usepackage{ae,aecompl}

\usepackage{graphicx}	% Including figure files
\usepackage{amsmath}	% Advanced maths commands
\usepackage{amssymb}	% Extra maths symbols

\title[The Detection of Dust around NN\,Ser]{The Detection of Dust around NN\,Ser}

\author[A. Hardy et al.]{Adam Hardy$^{1,2}$,\thanks{E-mail:adam.hardy@postgrado.uv.cl}
Matthias R. Schreiber$^{1,2}$,
Steven G. Parsons$^{1}$,
Claudio Caceres$^{1,2}$,
\newauthor Carolyn Brinkworth$^{3,4}$,
Dimitri Veras$^{5}$,
Boris T. G\"ansicke$^{5}$,
Thomas R. Marsh$^{5}$,
\newauthor Lucas Cieza$^{2,6}$
\\
$^{1}$Instituto de F\'isica y Astronom\'ia, Universidad de Valpara\'iso, Av. Gran Breta\~na 1111, Valpara\'iso, Chile\\
$^{2}$Millennium Nucleus ``Protoplanetary Disks in ALMA Early Science'', Universidad de Valpara\'iso, Casilla 36-D, Santiago, Chile\\
$^{3}$Spitzer Science Center, IPAC, Caltech, Pasadena, CA 91125\\
$^{4}$National Center for Atmospheric Research, Boulder, CO 80301\\
$^{5}$Department of Physics, University of Warwick, Coventry CV4 7AL, UK\\
$^{6}$N\'ucleo de Astronom\'ia de la Facultad de Ingenier\'ia, Universidad Diego Portales, Av. Ej\'ercito 441, Santiago, Chile\\
}

\date{Accepted XXX. Received YYY; in original form ZZZ}

\pubyear{2016}

\begin{document}
\label{firstpage}
\pagerange{\pageref{firstpage}--\pageref{lastpage}}
\maketitle

\begin{abstract}
Eclipse timing variations observed from the post common-envelope binary (PCEB) NN\,Ser offer strong evidence in favour of circumbinary planets existing around PCEBs. If real, these planets may be accompanied by a disc of dust. We here present the ALMA detection of flux at 1.3\,mm from NN\,Ser, which is likely due to thermal emission from a dust disc of mass $\sim0.8\pm0.2\,M_{\oplus}$. We performed simulations of the history of NN\,Ser to determine possible origins of this dust, and conclude that the most likely origin is, in fact, common-envelope material which was not expelled from the system and instead formed a circumbinary disc. These discs have been predicted by theory but previously remained undetected. While the presence of this dust does not prove the existence of planets around NN\,Ser, it adds credibility to the possibility of planets forming from common-envelope material in a `second-generation' scenario. 
\end{abstract}


%%%%%%%%%%%%%%%%%%%%%%%%%%%%%%%%%%%%%%%%%%%%%%%%%%

%%%%%%%%%%%%%%%%% BODY OF PAPER %%%%%%%%%%%%%%%%%%

\section{Introduction}

Post-common envelope binaries (PCEBs) are among the most peculiar binary stars in existence. They contain at least one compact object (either a white dwarf, neutron star or black hole) and a companion that orbits at very low separations. These separations are so low that the stars could not always have been in this configuration, as the compact object progenitor would have completely engulfed its companion when on its giant branch. Instead, the star's history was likely one of a main-sequence binary with separations of order 1\,au. The evolution of the more massive star onto the giant branch would have then caused dynamically unstable mass transfer onto its companion, resulting in a common-envelope (CE) of material. The CE surrounds the future compact object and the companion but due to
drag forces within the CE, orbital energy is extracted from
the binary causing the two stars to spiral inwards. When enough energy is transferred to the envelope, it will then be expelled, leaving the compact object and companion star that we can observe today \citep[e.g.][]{Paczynski1976,Webbink1984,Zorotovic2010}. 

Although there is general agreement that the scenario outlined above describes the basics of PCEB formation, many questions regarding PCEBs still remain. 
One such question which has gained considerable attention in recent years is what effect the violent evolution of these stars might have on any circumbinary material, including planets or brown dwarfs. This is particularly relevant, as many PCEBs display a potential signature of substellar objects in orbit: almost all eclipsing PCEBs display variations in their measured eclipse timings \citep{Zorotovic2013}, and these eclipse timing variations (ETVs) have been attributed to circumbinary objects periodically moving the center of mass of the binary system \citep[e.g.][]{Guinan2001}. If these ETVs are indeed due to circumbinary objects, the question emerges of whether these are first-generation objects which survived the common envelope evolution, or instead formed afterwards in a `second-generation' scenario \citep{Volschow2014,Schleicher2014}. Both scenarios face difficulties - the first-generation scenario has trouble in keeping progenitor planets in stable orbits throughout the common-envelope evolution \mbox{\citep{Mustill2013}}, 
implying that not many first-generation planets are expected to survive. The fact that ETVs have been detected around almost all eclipsing PCEBs instead suggests that these planets must commonly form from material remaining after envelope expulsion \citep{Zorotovic2013}. This second-generation scenario, however, faces its own challenges. For example, a study by \citet{Bear2014} found that forming second-generation planets around many PCEBs would require a very efficient planet-making process, in which more than 20\% of the disc material goes into planets.  

It is clear that there is some uncertainty surrounding how the potential planets around PCEBs formed, but there is also an ongoing discussion on whether these planets even exist.  Suggestions that the planetary interpretation is incorrect include the observation that the majority of planetary models used to explain the eclipse timing variations are unstable \citep[e.g.][]{Horner2012,Horner2013} or fail drastically when confronted with more recent eclipse timing measurements \citep{Parsons2010,Bours2014}. 
Further evidence comes from recent observations of the PCEB V471\,Tau, in which direct imaging was carried out to search for the predicted companion \citep{Hardy2015}. V471\,Tau was the first eclipsing PCEB discovered, and therefore has a long baseline of timing measurements clearly showing ETVs \citep{Nelson1970,Lohsen1974}. These data allowed accurate prediction of the brown dwarf's brightness and separation from the PCEB, but despite these parameters being within the capability of the direct imaging observations, no brown dwarf was detected. It is therefore highly likely that another mechanism is causing the ETVs seen in V471\,Tau. The exact nature of this mechanism is uncertain, but one possibility in the case of V471\,Tau is the Applegate mechanism, in which the ETVs are prescribed to periodic changes in the magnetic field of the main-sequence companion \citep{Applegate1992}. 

Whilst non-planetary explanations for eclipse timing variations in PCEBs have been gaining some ground, one PCEB which so far remains robust against the above criticisms is NN\,Ser. NN\,Ser is a relatively young PCEB with a white dwarf age of $\sim$1.3\,Myr \citep{Schreiber2003}. The companion to the white dwarf is an M4 type star orbiting with a period of 0.13\,days \citep{Brinkworth2006, Parsons2010a}, and NN\,Ser displays eclipse timing variations that are well fit by 2 planetary mass bodies \citep{Beuermann2010}. Unlike many other PCEBs, this planetary model has correctly predicted more recent timing measurements, and no other mechanism has yet been proposed to explain its behaviour \citep{Beuermann2013,Marsh2014,Parsons2014}. As such, NN\,Ser is perhaps the best PCEB with which to further test the planetary hypothesis. Its youth means that, if the second generation planetary scenario is correct, it is possible that NN\,Ser still possess protoplanetary disc material. If this protoplanetary disc material is already dissipated, or if the first generation formation scenario is correct, it is further possible that the planets would be present alongside a debris disc as observed in many systems.

To test this hypothesis, we searched for dust remaining after the common envelope around NN\,Ser. We used SOFI and Spitzer to search for hot dust close to the star, and ALMA to probe for cool dust farther out.
While we detect no conclusive excess emission at IR wavelengths, we clearly detect excess emission from NN Ser with ALMA which is likely due to thermal emission from a belt of cold dust. We conclude that this dust is likely a circumbinary disc, formed of material left over from the CE. 

\section{Observations}\label{obs}
 
In close binaries such as NN\,Ser, the main-sequence star is prone to reflection effects from the white dwarf, resulting in an excess above the stellar photosphere. This reflection effect, coupled with the systems eclipsing nature, means that the emission of NN\,Ser is phase-dependant. 
As a result, when searching for excess emission above the stellar photosphere, it is crucial that all data be taken at the same phase. In our observations, we chose this phase to be just after the end of the white dwarf eclipse by the main-sequence star. This phase was chosen as the heated face of the main-sequence star will be pointed away from us, minimising the reflection effect.
In addition to the new data presented here, we also took the optical data from \cite{Parsons2010a}, and calculated the emission at our chosen phase to give us consistent data at shorter wavelengths. 

\subsection{SOFI Observations}
$J, H$ and $K$ band observations of NN\,Ser were obtained with the Son of Isaac instrument (SOFI) \citep{Moorwood1998} mounted on the New Technology Telescope (NTT) at La Silla observatory, Chile in 2010-04. The observations were made in fast photometry mode and covered almost an entire binary orbit in the $J$-band, half an orbit in the $H$-band and the eclipse of the white dwarf in the $K$s-band. We windowed the detector to achieve a cycle time of 10 seconds and offset the telescope every 10 minutes in order to improve sky subtraction. The $J$-band observations were slightly affected by thin clouds.

The dark current removal and flat-fielding were performed in the standard way. Sky subtraction was achieved by using observations of the sky when the target had been offset. The average sky level was then added back so that we could determine the source flux and its uncertainty with standard aperture photometry, using a variable aperture, within the ULTRACAM pipeline \citep{Dhillon2007}. A comparison star was used to account for variations in observing conditions and to flux calibrate the data by using its 2MASS magnitudes \citep{Skrutskie2006}.

\subsection{Spitzer Observations}
The Spitzer data were obtained during Cycle 3, as part of program 30070. Data were taken with the Infrared Array Camera \citep[IRAC; see][]{Fazio2004} with 25 x 100s dithers in all 4 channels (3.6 um - 8.0 um). The data reduction for all 4 channels was carried out on the Corrected Basic Calibrated Data frames (CBCDs) downloaded from the Spitzer archive. The CBCDs were overlap-corrected and combined with the standard Spitzer Science Center (SSC) software MOsaic and Point Source EXtraction \citep[MOPEX;][]{Makovoz2006}, using dual outlier rejection, to create a single mosaicked image for each channel. Rejected frames were noted, and the mosaic was used to identify the position of the target. No further analysis was carried out on the mosaics. 

The original downloaded CBCDs were corrected for array location dependence using the correction frames provided by the SSC. Aperture photometry was carried out on the corrected CBCD frames using an aperture radius of 3 pixels and a sky subtraction annulus from 12-20 pixels. 

The photometry was converted from MJy sr$^{-1}$ to mJy and aperture corrected using the standard aperture corrections provided by the SSC. No pixel phase correction was applied due to the averaging effect of the dither pattern. Additionally, no colour correction was applied since we quote the isophotal wavelengths, thereby reducing the color dependency of the flux calibration to negligible within our uncertainties. 

We rejected photometry from CBCDs that were flagged by the MOPEX Dual Outlier Rejection algorithm during the initial mosaic process. The quoted flux densities for each channel are the unweighted mean of the photometry from the remaining frames. The uncertainties were estimated from the rms scatter on the photometry from the individual frames, divided by the square root of the number of frames. The quoted uncertainties are either our calculated uncertainty or the IRAC instrument calibration uncertainty \citep{Reach2005}, whichever was larger.
 
The Spitzer observations did not cover the orbital phase used in the optical and NIR observations. We therefore used the light curve model from \cite{Parsons2010a} and adjusted the output wavelength to match the Spitzer bands. We then fitted the Spitzer data, keeping all the parameters fixed at the same values as in \citet{Parsons2010a} except the temperature of the M star, which we allowed to vary in order to fit the amplitude of the reflection effect. We used this model to determine the brightness of the system in the Spitzer bands at the correct phase. 
\\
\subsection{ALMA Observations}
The Atacama Large Millimeter/submillimeter Array (ALMA) observations were conducted on 2014-04-30 and repeated on the 2015-01-22, as the rms sensitivity of the data did not reach the requested value of 25$\mu$J in the first instance. As no flux was detected in the first observation, we were unable to accurately account for the ALMA pointing error and line-up the different data sets. We therefore focus on the second measurement set only. The single continuum mode in Band 6 was used, implying a total bandwidth of 7.5 GHz with individual channel widths of 15.625\,MHz. 
39 antennas were used, with minimum and maximum baselines of 15.1\,m and 348.5\,m respectively.

The calibration sources associated with these observations
were J1337-1257 for band-pass calibration, and J1550+0527
for phase calibrations. The observations consisted of 5 scans of 6.87 min each, translating to a total time on the science target of 34.35 min.

\begin{figure}
	\includegraphics[width=\columnwidth]{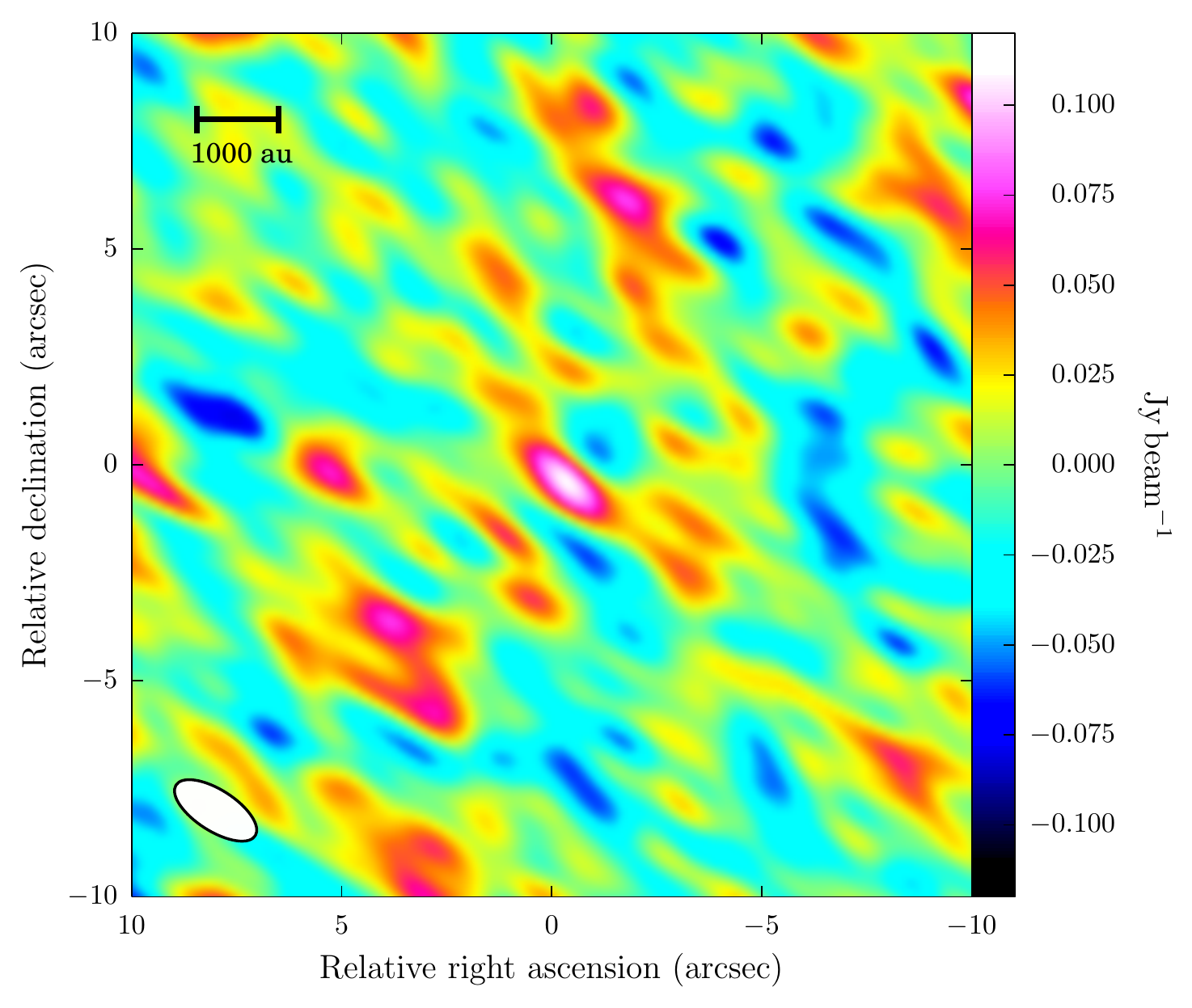}
    \caption{ALMA image of NN\,Ser at 1.3\,mm generated with the CLEAN algorithm using natural weighting. The emission is unresolved, but the beam size (the white ellipse) confines the emission to within 1000\,au of NN\,Ser}
    \label{fig:alma}
\end{figure}

Standard calibration steps were applied to the data, and the resulting visibilities were deconvolved using the CLEAN algorithm with natural weighting to create the final image (Fig \ref{fig:alma}). To obtain the total flux at 1.3mm, a point-source fit to the visibilities was performed using the uvmodelfit task in casapy version 4.2.2, resulting in a flux value of 0.11$\pm$0.03\,mJy. As the emission at this wavelength is likely due to thermal emission from dust at large orbital radii (see section \ref{thermal}), the orbital phase of the PCEB should have no effect and was not taken into account.

\subsection{SED}
\begin{figure}
	\includegraphics[width=\columnwidth]{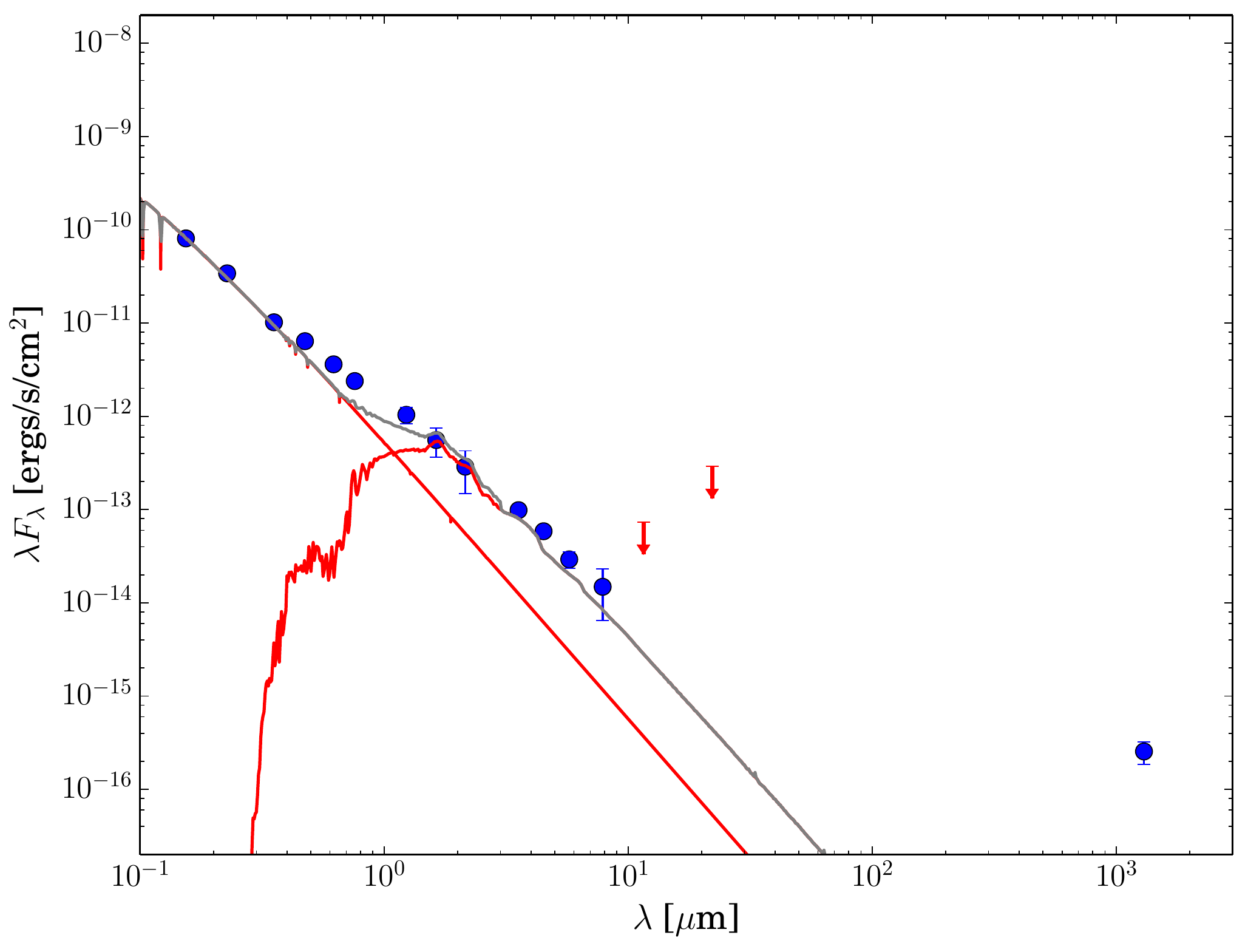}
    \caption{Spectral energy distributional of NN\,Ser at the end of the white dwarf eclipse. The red lines denote fluxes from a model 60,000\,K white dwarf and M4 main-sequence star, with the grey lines as their sum. The blue points are the data, and the red arrows are upper limits from WISE. There is a marginal excess above the stellar photosphere in the range 0.47-0.75\,$\mu$m, but this may be the result of heating of the main-sequence star by the white dwarf. An excess at 1300\,$\mu$m is clearly detected however.}
    \label{fig:SED}

\end{figure}

The SED of NN\,Ser (table \ref{tab:SED}) is shown in figure \ref{fig:SED}, along with models of a 60,000\,K white dwarf \citep{Koester2010} and a M4 type companion \citep{Allard2012}. Although there are some hints of an excess in the range 0.621-1.23 $\mu$m, this may well be the effect of a slightly non-uniform temperature across the main-sequence star caused by heating from the white dwarf. As such, there is no conclusive excess emission at wavelengths less than 8$\mu$m. In the ALMA band on the other hand, there is a clear excess. 

\begin{table}

	\centering
	\caption{SED data for NN\,Ser. (1) Parsons et al. (2010a). (2) This work.}
	\label{tab:SED}
	\begin{tabular}{cccc} % four columns, alignment for each
		\hline
		Wavelength($\mu$m) & Flux(mJy) & $\sigma_{Flux}$ & Ref\\
		\hline
	0.154 &  4.16 & 	0.01 & (1)\\
	0.227 & 	2.59	 &  0.01 & (1)\\
	0.354 &  1.20 &  0.01 & (1)\\
	0.474 &  1.011 &  0.003 & (1)\\
	0.621 &  0.748 &  0.004 & (1)\\
	0.758  & 0.604 &  0.004 & (1)\\
	1.23 &  0.43 &  0.09 & (2)\\
	1.64 &  0.3 &  0.1 & (2)\\
	2.15 &  0.2 &  0.1 & (2)\\
	3.56 &  0.117 &  0.005 & (2)\\
	4.50 &  0.088 &  0.005 & (2)\\
	5.74 &  0.056 &  0.011 & (2)\\
	7.87 &  0.039 &  0.022 & (2)\\
	1300 &  0.11 &  0.03 & (2)\\
		\hline
	\end{tabular}
\end{table}

\begin{figure*}
	\includegraphics[width=\textwidth]{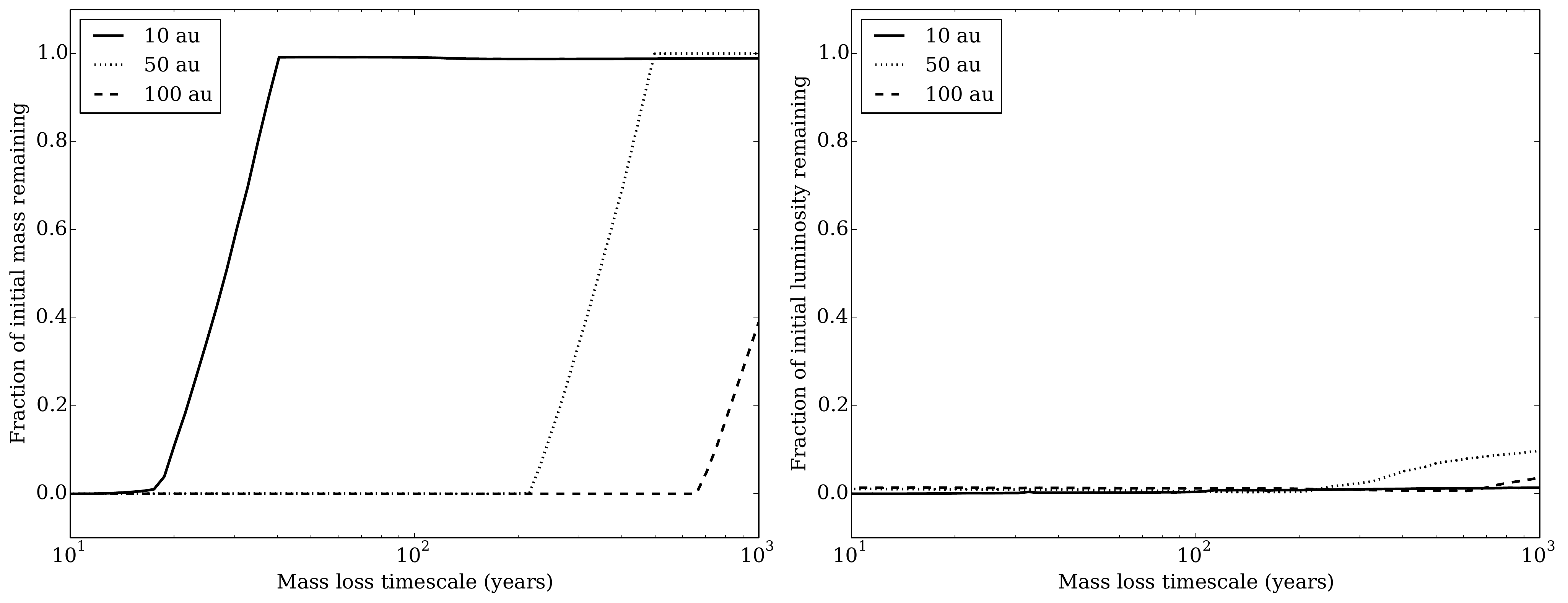}
    \caption{\textit{Left:} Fraction of the initial disc mass that remains after common-envelope evolution as a function of mass loss-timescale. Disc material is lost through a combination of radiation forces, stellar wind drag and the reduction in gravitational force due changing central mass. The three black lines reflect different initial radii for the debris disc material. \textit{Right:} Fraction of the initial disc luminosity which remains after the common-envelope evolution following the same processes as the figure to the left. The much lower fractions reflects the fact that the CE evolution more efficiently removes small particles, which dominate the surface area of the emitting disc.}
    \label{fig:ml_lost}
\end{figure*}

\section{Possible origins of the ALMA flux}

\subsection{Gyrosynchrotron emission}

Two plausible explanations exist for the excess emission detected with ALMA, with the first being gyrosynchrotron emission from material in the magnetic field between the white dwarf and the main-sequence star. Indeed, radio emission from the PCEB V471\,Tau has been detected and attributed to gyrosynchrotron emission, making a similar process plausible in NN\,Ser. However, the emission from V471\,Tau has been measured as $\sim3$\,mJy at 5GHz \citep{Patterson1993} and, if placed at the distance of NN\,Ser, this emission would be reduced to $\sim0.03$\,mJy. Even if it is assumed that the gyrosynchrotron emission follows a flat distribution all the way to the ALMA frequency of 230\,GHz, this level of emission would not explain our ALMA detection.  If gyrosynchrotron were truly causing this detection therefore, the emission from NN\,Ser would have to be several orders of magnitude stronger than for V471\,Tau. 
This is not likely to be the case, although this gyrosynchrotron emission cannot be conclusively ruled out without further observations. The spectral slope of gyrosynchrotron emission is expected to be very distinct, meaning that observation at just one nearby wavelength will easily resolve this issue.

\subsection{Thermal emission from dust}\label{thermal}

Alternatively, the emission may originate from a ring of dust around NN\,Ser. With just one SED point, the grain size distribution and surface density distribution of this disc cannot be constrained. However, the lack of clear IR excess suggests that the disc does not extend too close to the central binary and the ALMA beam size confines this emission to within 1000\,au of NN\,Ser. Furthermore, the dust emission at this wavelength is optically thin, meaning the dust mass can be estimated by a simple equation of the form $M_{\rm dust}=C_{\nu} \times F_{\nu}$ where $ C_{\nu}$ is a constant for a given frequency $F_{\nu}$ \citep{Andrews2005}.

We adopt the constant derived for 1.3mm by \cite{Cieza2008a}, and use the equation
\begin{equation}   
      M_{\mathrm{dust}}=0.566 \times \left[ \dfrac{F_{\nu}(1300)}{\rm mJy} \left( \dfrac{d}{140\,\rm pc} \right)^{2} \right] M_{\oplus}
\end{equation} 
to a derive value of $0.8\pm0.2\,M_{\oplus}$ for the dust mass around NN\,Ser, where a distance of $512\pm43$\,pc was assumed \citep{Parsons2010a}.
This approach contains several assumptions about the grain properties so this value should be treated with some caution. However, it allows comparison with discs around other objects, and we find that the dust mass obtained this way is similar to that of young debris disks around low-mas stars \citep[e.g.][and references therein]{Hardy2015a}.  

\section{Possible origins for the dust}\label{sims}

We see three possible origins for the dust. First, the dust may be debris disc material which existed before the CE and survived the evolution of the central binary. Second, the material we observe may have existed in larger planetesimals before the CE, which then collided as a result of the binary evolution to form dust. Third, the dust may have formed in the CE itself but was not successfully ejected from the system, and instead created a circumbinary disc. The violent evolution of the host binary may preclude some of these origin scenarios and to evaluate how realistic these scenarios are, we performed different simulations for dust production and survival around NN\,Ser. Three simulations were performed using a modified version of the N-body simulator MERCURY \citep{Chambers1999}. The modifications account for the additional forces felt by particles surrounding the binary.

During the CE, particles will be affected by radiation forces from the central AGB star, combined with the stellar-wind drag and a changing central mass associated with CE expulsion. We also account for radiation forces in the white dwarf-main sequence state and drag caused by the stellar wind from the M star. 
The luminosity of the system, and therefore radiation forces, will change as the star evolves however, and we therefore further modify the code to accept a time-varying luminosity input. This input was calculated using the Binary Star Evolution (BSE) code \citep{Hurley2002} with a white dwarf progenitor mass of 2.08\,$M_{\odot}$, companion mass of 0.11\,$M_{\odot}$, eccentricity of 0, CE efficiency of 0.25, metallicity of 0.02 and a radii at the start of Roche Lobe overflow of 194\,$R_{\odot}$ \citep{Beuermann2010}. All other BSE parameters were kept at their default. The simulations involved a total mass loss of 1.545\,$M_{\odot}$ to bring the white dwarf mass down to the measured value of 0.535\,$M_{\odot}$. This mass ejection was modelled as being ejected preferentially in the binary plane causing a drag for any disc material. For further details, see appendix \ref{appendix}.

\subsection{Surviving first-generation debris disc}

Debris discs have been found around main-sequence binary stars with similar separations and ages to that of the NN\,Ser progenitor \citep{Trilling2007}. It could therefore be that the dust we detect is a similar circumbinary debris disc that survived the CE. However, the strong radiation forces during the AGB and CE phases may remove all the small dust which dominates the luminosity in such discs. 

To determine if debris disc material can survive the CE, we run simulations of a debris disc located at $r=$ 10, 50 and 100\,au, with the width of the disc $dr=r$/2. The minimum grain size in this disc was set at the blowout radius before the AGB commenced  (i.e. 12$\mu$m), and the maximum was set at 60\,km as this value had previously been found as the largest that contributes to a collisional cascade \citep{Wyatt2007}. The grain size distribution was set to follow the standard relation of particles in a collisional cascade, $n(D)= KD^{2-3q}$
, with $q$=11/6. \citep{Tanaka1996}. We evolve this disc through 4.5\,Myr of the AGB, followed by a CE with an associated mass-loss timescale. Finally, the simulations are continued for 1.3\,Myr with a white dwarf-main sequence binary. 

The timescale over which the mass loss takes place in the CE is not a well-determined parameter, as it depends on the unknown efficiency with which the CE material can extract angular momentum from the binary. We therefore leave it as a free parameter in our simulations. However, this timescale must be shorter than the thermal timescale of the envelope otherwise the extracted orbital energy would be radiated away and no envelope expulsion would take place. This places a limit on the CE phase of $\sim$ 10$^{3}$ years \citep{Webbink1984} and we therefore perform simulations with mass-loss timescales of between 1 and 10$^{3}$ years.

In our simulations, we assume initially that the particles do not interact (and there are therefore no collisions) to identify what proportion of material that may have existed before the AGB is lost, and how this lost material reflects on the change in luminosity of the disc. The fractional initial disc mass which is lost as a result of CE evolution is displayed in figure \ref{fig:ml_lost}, left panel. It is apparent that for rapid mass loss (less than $\sim$40 years), almost all material is lost in all simulations, whereas a longer mass-loss timescale can allow almost all mass to survive in the discs with initial radii of 10 and 50\,au. The left panel of figure \ref{fig:ml_lost} is somewhat deceptive however, as the modelled grain size distribution places the majority of the mass in larger particles, whereas the surface area (and therefore flux) is dominated by smaller particles that are more easily lost by radiation pressure and drag. This dominance over surface area that the small grains possess means that discs which retain the bulk of their mass in large bodies may still emit considerably less at 1.3\,mm. We therefore plot the fractional change in 1.3\,mm flux that one would expect as a result of the CE evolution in figure \ref{fig:ml_lost}, right panel. From this panel, it becomes apparent that any dust which existed before the CE will likely be lost, reducing the luminosity of the disc considerably. The most favourable configuration for luminosity still suffers a $\sim$90\% decrease in flux, suggesting that if the dust detected was indeed comprised of material that survived the CE, than the progenitor disc would need a dust mass that is unrealistically high ($\sim 8\,M_{\oplus}$) when compared to discs of a similar age \citep[e.g.][]{Panic2013,Hardy2015a}. A high dust-mass disc existing around the NN\,Ser progentitor is made more implausible by the observations that fractional luminosity of debris discs appears to decrease with age \citep{Rieke2005, Su2006}
, and that relatively few discs are observed around M-type stars older than 10\,Myr, likely due to their increased stellar wind drag \citep{Plavchan2005}. It is therefore very unlikely that the dust observed existed before the CE evolution.

\subsection{Second-generation debris disc}

As debris discs are replenished by collisions, the above simulations, in which the particles do not interact, perhaps underestimate the amount of small grains that exist after CE evolution.
In fact, the rapid mass-loss which is associated with the CE expulsion can induce large eccentricities in the orbiting bodies \citep[e.g.][]{Veras2011}, which may cause considerable collisions between planetesimals. We therefore repeat the simulations described, but this time only simulate the largest planetesimals and track their collisions. The number of collisions in these simulations will naturally depend on the number of planetesimals present and therefore the total mass of the disc. We find that no collisions occur, but our simulations are limited to low mass discs, as modelling $\ge$20000 planetesimals with MERCURY is computationally challenging. To estimate the number of collisions for higher mass discs, we instead calculate the collisional timescale of the planetesimals which remain after the CE has been expelled. We use the prescription of \cite{Wyatt2010} to calculate the collisional rate of the largest planetesimals in our simulation, $R_{\rm c}$:
\begin{equation}   
      R_{\rm c}=K M_{\rm tot}\,\upsilon_{\rm k}^{8/3}a^{-3}(4\pi I_{\rm max})^{-1}\left[ 0.54e^{5/3}(1-e^{2})^{-4/3}\right]     
\end{equation} 
with 
\begin{equation}   
K=9.5\times10^{-6}\rho^{-1}D_{\rm max}^{-1}\,Q_{D}^{-5/6}     
\end{equation}

where $\rho$ is the density of the planetesimals, $D_{\rm max}$ the radii of the largest planetesimal in the collisional cascade in km, $M_{\rm tot}$ is the total mass of the disc in $M_{\oplus}$, $\upsilon_{\rm k}$ the Keplerian velocity of the planetesimals, $a$ their semi-major axis, $I_{\rm max}$ their maximum inclination, $e$ their eccentricity and $Q_{D}$ their planetesimal strength in Jkg$^{-1}$. The value of $Q_{D}$ varies as a function of planetesimal radius and composition \citep[e.g.][]{Krivov2005}, and we adopt a conservative value of $1 \times 10^{4}$\,Jkg$^{-1}$ for rocky planetesimals of radius 60\,km. The collisional rate from this equation was then inverted to calculate the collisional lifetime.  

\begin{figure}
	\includegraphics[width=\columnwidth]{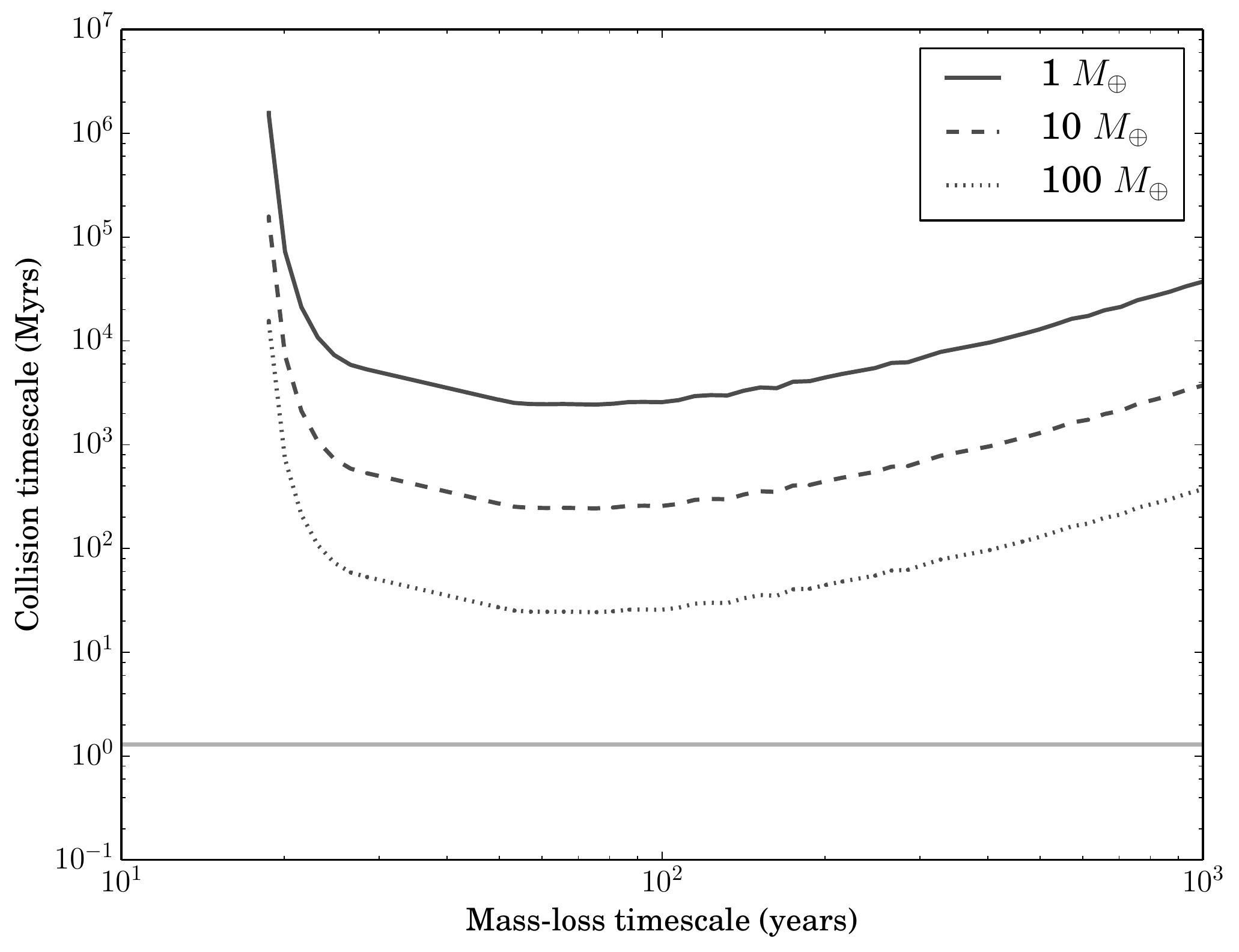}
    \caption{Collisional timescale of a planetesimal belt initially located at 10\,au, following CE evolution of differing mass loss timescales. The collisional timescale depends on total disc mass, and even $\sim$100\,$M_{\oplus}$ of material will not cause the collisional timescale to drop below the age of NN\,Ser (solid horizontal line). It is therefore unlikely that collisions can create the dust seen around NN\,Ser.}
    \label{fig:tcoll}
\end{figure}

For simulations of the 50 and 100 au discs, the large semi-major axis obtained by the planetesimals causes collision timescales $\geq 10 ^{4}$\,Myr even for disc masses as high as 100\,$M_{\oplus}$. For the disc initially at 10\,au, the collision timescale is somewhat lower, but a large disc mass of $\sim$100\,$M_{\oplus}$ will still only experience frequent collisions on timescales $\gg 20$\,Myr (see Fig. \ref{fig:tcoll}). The dust in these discs will therefore not be replenished within the 1.3\,Myr age of NN\,Ser. This result is similar to that of simulations of planetessimals around single AGBs, which found that the collisional timescale can increase up to the Hubble time in some cases \citep{Bonsor2010}, resulting in constant debris disc masses in the white dwarf phase. 

Although the collisional timescale in both our simulations and those around a single white dwarf are extremely long, it has been suggested that the presence of a planet on an unstable orbit can rejuvenate the collisions between planetesimals and create a debris disc \citep{Debes2002, Dong2010}. Indeed, a second-generation debris disc which has been attributed to the influence of a planet has potentially been observed around the central star of the helix nebula \citep{Su2007,Bilikova2012}. If the ETVs in NN Ser were due to first-generation planets, a debris disk caused by this mechanism remains a possibility. However, a major problem with this scenario is that progenitor systems for the first-generation planets around NN\,Ser have been studied in detail and found to be unstable over the main-sequence lifetime \citep{Mustill2013}. Therefore, given the increased collisional lifetime as a result of the CE and the difficulties in maintaining a first-generation planetary system to cause collisions, our results points toward the observed dust being of a different origin.

\subsection{Remaining common-envelope material}

A further possibility for the origin of the dust detected with ALMA, is that it is `second-generation' material, left over from the common envelope itself. It has been observed that the AGB phase can create significant amounts of dust \citep{Hoogzaad2002,Lebzelter2006}, and models specifically of CE evolution likewise suggest that dust formation can be very efficient in this environment \citep{Lu2013}. The evolutionary history of PCEBs then implies that this dust carries a certain amount of angular momentum, which facilitates the formation of a disc of CE material. SPH simulations have suggested that a large amount of the material lost from the primary can, in fact, remain bound to the system \citep{Sandquist2000,Ricker2012,Passy2012}. Furthermore, dusty discs resulting from the AGB phase have been detected around both post-AGB binaries \citep[e.g.][]{vanwinckel2009} and single neutron stars \citep{Wang2006}. However, models suggest the dust produced during the AGB will have radii of $\sim0.3\mu$m \citep{Yasuda2012}, and grains of this size will also be affected by radiation forces from the newly formed hot white dwarf. These grains therefore may not survive long enough to be observed around the 1.3\,Myr old PCEB. 

\begin{figure}
	\includegraphics[width=\columnwidth]{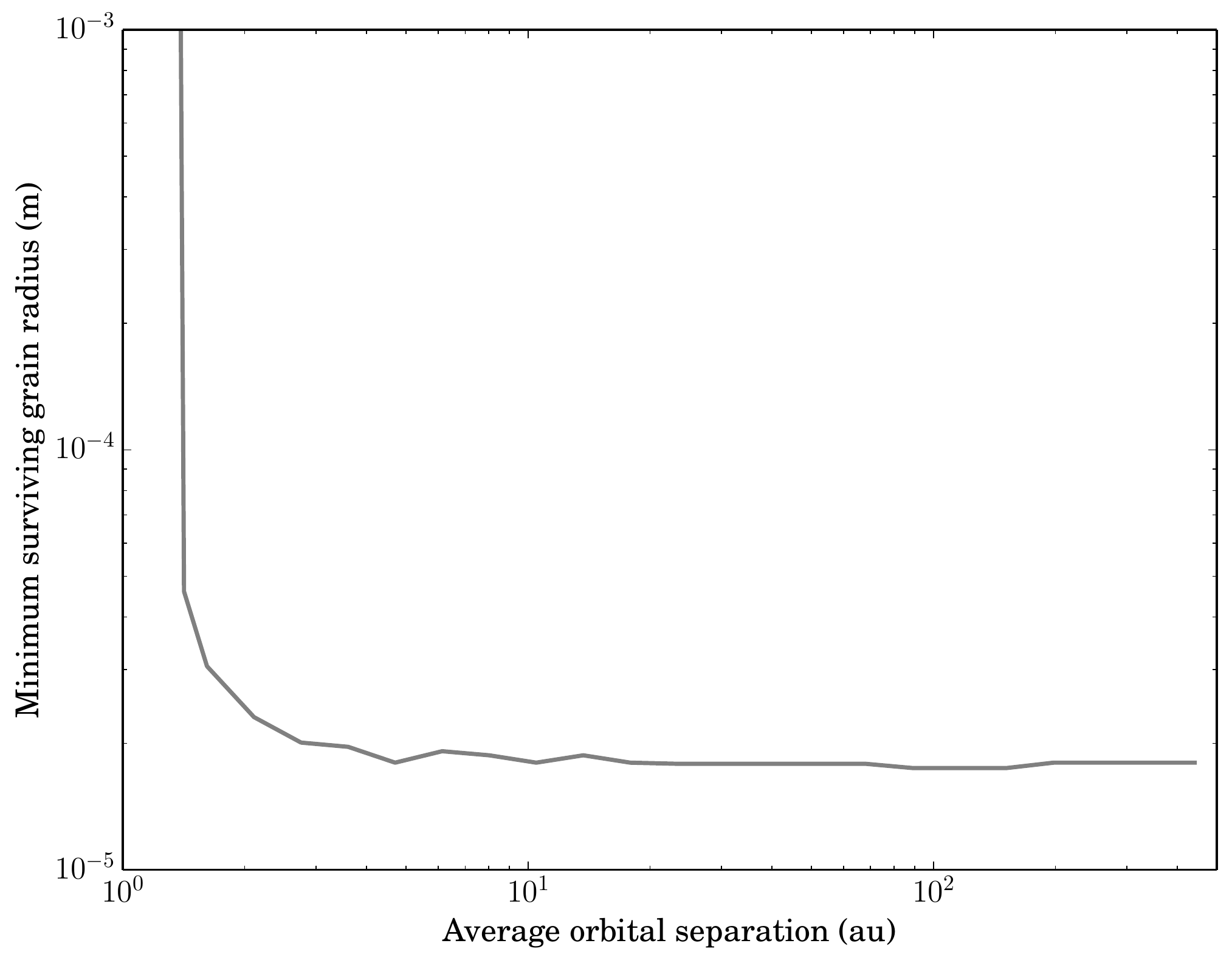}
    \caption{Radius of the smallest grain that can survive the 1.3\,Myr that has passed since the CE of NN\,Ser whilst experiencing radiation forces and stellar wind drag. This is relevant to dust which was created in the CE and remained in the system as part of a circumbinary disc. These simulations do not include grain growth/reprocessing, which could allow smaller particles to be present in the disc.}
    \label{fig:fallback}
\end{figure}

To investigate the size of grains which may survive the PCEB phase, we again run simulations using our modified version of MERCURY for a population of grains with sizes ranging from 0.01$\mu$m to 1$\,$cm. This simulation is commenced with the binary in its PCEB configuration, so dust is only prone to radiation forces from the white dwarf-main sequence binary and a small drag force due to a stellar wind from the main-sequence star. In this case, the grains size distribution is unknown, so we instead simulate a uniform number distribution of grains distributed logarithmically between 0.5\,au and 500\,au. We then record the average orbital radii for each particle that survives the simulation, and plot the minimum value of the particle radii in figure \ref{fig:fallback}.

Grains smaller than than $\sim20\mu$m are removed in our simulations, and this is considerably larger than the grain size believed to be generated within the AGB of $\sim0.3\mu$m \citep{Yasuda2012}. One might therefore expect grains made within the CE to be instantly lost. However, if the amount of gas and dust is large enough, the disc may be optically thick and have an inner region where radiation pressure is negligible \citep{Takeuchi2003, Olofsson2009}. In these regions, its conceivable that grains could grow to 100$\mu$m in as little as $\sim$10$^{4}$ years \citep{Dullemond2005}, in the same manner as in protoplanetary discs. Models of the CE phase differ in their estimates of how much material can remain bound to the system, with values ranging from 10\% \citep{Kashi2011} to as much as 97\% of the CE \citep{Passy2012}. In either case, this would suggest a large amount of gas ($\gtrsim$0.15$M_{\odot}$ in the case of NN\,Ser) remains bound in a disc, so the existence of optically thick regions is certainly plausible. This gas might still exist alongside the detected dust, allowing one observational avenue by which this scenario can be tested.  

Dust attributed to the AGB has already been detected around several post-AGB binary stars in the near and mid-IR \citep{deRuyter2006,vanwinckel2009,Hillen2016}.
The post-AGB nature of these stars naturally means they are much younger than NN\,Ser, but studies of these objects suggests that already the grains have been through considerable reprocessing with grain sizes larger than $\sim$2$\mu$m \citep{Gielen2008}, adding credibility to the possibility that this dust is left over AGB material. It has also been  pointed out that there is some similarity between these discs and protoplanetary discs found around young stellar objects \citep{deRuyter2006}, opening the possibility that discs around these evolved binaries may provide a new avenue with which to study protoplanetary disc evolution and dissipation. 

\section{Planets}\label{discuss}

Given the convincing planetary fit for NN\,Ser, an obvious question is what the detection of this dust implies for the putative planets around PCEBs. 
Our simulations suggest that the detected disc is most-likely left over from the CE. The theory that the planets around NN\,Ser formed in a second-generation scenario \citep{Volschow2014} may gain some support from this detection, as the material necessary to form these planets seems to exist. The dust detected might then be material which simply has not efficiently grown yet, or it could be material that has grown, but since been involved in collisions. The growth of planets $\geq$1000\,km in radii would stir up collisions between smaller bodies and might rejuvenate the population of small dust, as described by models of `self-stirring' \citep{Kenyon2004}. The timescales required to achieve this planet formation and subsequent self-stirring are also consistent with the 1.3\,Myr age of NN\,Ser. However, although second generation planets gain some support from our results, the detection of dust does not directly correlate with the existence of planets around NN\,Ser, as the dust may simply be CE material that has not yet grown into larger bodies or been ejected. Furthermore, self-stirring models can form small planets capable of stirring the disc within 1.3\,Myr, but the planets predicted from the ETVs of NN\,Ser have masses of 6.97\,\Mjup and 1.73\,\Mjup \citep{Beuermann2013}. It remains uncertain if the formation of these much more massive planets can occur within the 1.3\,Myr age of NN\,Ser. As discussed in detail in \citet{Zorotovic2013}, the disc instability model can form planets within this timescale, but only at orbital separations much further than those predicted by the ETVs. Significant orbital migration would therefore need to have occurred if this were the case. Classical core accretion on the other hand would be able to form planets at their predicted location, but would struggle to form planets within 1.3\,Myr. For example, models which estimate the formation timescale of Jupiter within the solar system suggest that gas giants gain the majority of their mass at times $\gtrsim$2.5\,Myr \citep{Lissauer2009}.   

Another potential implication that this detection has on the field of planets around PCEBs is that these circumbinary discs can perhaps interact with the binary itself \citep{Artymowicz1991, Kashi2011}, causing the binary to lose angular momentum and slowly change the period of its orbit. As the planets around PCEBs are inferred from fits to the eclipse timing variations, this extra angular momentum loss would need to be accounted for. 
Before the planetary fits of NN\,Ser gained credibility, the possibility of a circumbinary disc being the cause of the perceived change in angular momentum was investigated by \cite{Chen2009}. Their model relied on a disc which obtained its mass through stellar wind from the M star, and they concluded that the disc was not massive enough to cause strong angular momentum loss unless a large fraction of the wind can end up in the disc ($\sim$10\%) and the wind loss rate is ultra-high ($\sim10^{-10}\,$\Msun yr$^{-1}$). If the disc observed here is comprised of common-envelope material however, then the total mass (including gas) might be very large, making it massive enough to extract angular momentum without the need of a high wind loss rate. 

Models of angular momentum loss due to circumbinary discs around cataclysmic variables are well developed \citep{Spruit2001,Taam2001}, but they rely on some knowledge of the surface density and inner radius of the disc. If the disc around NN\,Ser follows the surface density profile of a viscous accretion disc, then the surface density might only reach significant levels at small orbital radii. At such radii, one would expect a strong IR excess in the SED which is not seen, perhaps arguing against this affect being significant. Nonetheless, the effect of this circumbinary disc on the central binary offers interesting opportunities for future work.

\section{Conclusions}\label{conclude}

We present the detection of 1.3\,mm flux from the PCEB NN\,Ser, and find most plausible explanation for this flux is thermal emission from a circumbinary disc. We run simulations of the history of NN\,Ser to investigate what material can survive the extra forces associated with the AGB and common-envelope. Given the difficulties in creating/maintaining a debris disc of primordial material in our simulations, we find that a disc of left over common-envelope material is the most likely explanation for the detection. Such discs are predicted by theory, but have not previously been observed around a PCEB. This detection therefore adds credibility to the theory that second-generation planets might exist around NN\,Ser.

\section*{Acknowledgements}

We would like to thank the anonymous referee for their time and helpful comments. In addition, AH, MRS, CC and LC acknowledge support from the Millennium Nucleus RC130007 (Chilean Ministry of Economy). MRS, CC, SP and LC also acknowledge support from FONDECYT (grants 1141269, 3140592, 3140585 and 1140109). DV and BTG have received funding from the  European Research Council under the European Union's Seventh Framework Programme (FP/2007-2013)/ERC Grant Agreement n. 320964 (WDTracer). TRM was supported by STFC grant \#ST/L000733. Based on observations made with ESO Telescopes at the La Silla Paranal Observatory under programme ID 085.D-0541. This work is based, in part, on observations made with the Spitzer Space Telescope, which is operated by the Jet Propulsion Laboratory, California Institute of Technology under a contract with NASA. This paper makes use of the following ALMA data: ADS/JAO.ALMA\#2013.1.01342.S. ALMA is a partnership of ESO (representing its member states), NSF (USA) and NINS (Japan), together with NRC (Canada) and NSC and ASIAA (Taiwan) and KASI (Republic of Korea), in cooperation with the Republic of Chile. The Joint ALMA Observatory is operated by ESO, AUI/NRAO and NAOJ.

%%%%%%%%%%%%%%%%%%%%%%%%%%%%%%%%%%%%%%%%%%%%%%%%%%



\bibliographystyle{mnras}
\bibliography{PCEBs} % if your bibtex file is called example.bib

%\begin{thebibliography}{99}
%\end{thebibliography}

%%%%%%%%%%%%%%%%%%%%%%%%%%%%%%%%%%%%%%%%%%%%%%%%%%

%%%%%%%%%%%%%%%%% APPENDICES %%%%%%%%%%%%%%%%%%%%%

\appendix

\section{Simulation details}\label{appendix}

Simulations were performed using a modified version of the N-bdy code MERCURY \citep{Chambers1999}. These modifications allowed the inclusion of Radiation forces, stellar mass loss and stellar wind drag, the details of which are described below.  

\subsection*{Radiation forces}

At each timestep in the MERCURY simulator, each particle in the simulation felt an additional acceleration of the form \citep{Burns1979}:
\begin{equation}   
accl=\dfrac{GM_{\ast}\beta}{r^{2}}\left( \boldsymbol{\hat{r}}-\dfrac{v_{r}\boldsymbol{\hat{r}}+v}{c}\right) 
\end{equation} 
where $G$ is the gravitational constant, $M_{\ast}$ the mass of the central object, $\beta$ the ratio of the force of radiation pressure to gravity, $r$ the orbital separation of the particle, $\boldsymbol{\hat{r}}$ the unit vector corresponding to $r$, $c$ is the speed of light and $v_{r}=v\boldsymbol{\hat{v}}$ accounts for the Dopplar shift in the radiation seen by the particle. The value of $\beta$ was calculated using the equation 
\begin{equation}   
\beta =\dfrac{3L_{\ast}Q_{pr}}{16\pi cGM_{\ast}\rho R}
\end{equation} 
where $L_{\ast}$ is the luminosity of the star, $\rho$ the density or particles, $R$ their radius and $Q_{pr}$ the radiation pressure efficiency which was set at 1. 

\subsection*{Stellar mass loss}

The exact behaviour of how mass is lost during CE evolution is not known. We therefore assume that mass loss occurs only in the time described by the `mass-loss timescale' parameter (i.e. no mass loss occurred in the preceding RGB and AGB phases). We further assume that during this period the mass is lost linearly. We use the Bulirsch-Stoer integrator in MERCURY with a variable time-step, and to improve the accuracy of our simulations we apply this mass-loss as each sub-step of the Bulirsch-Stoer algorithm \citep[e.g.][]{Veras2013}. 

\subsection*{Stellar wind drag}

The accelerations on the particles due to stellar wind drag were calculated using the equations from \cite{Garaud2004}, expressed in the notataion of \cite{Veras2015} as:
\begin{equation}   
a_{swd}=\begin{cases}
    \left(\dfrac{\rho_{g}v_{g}}{\rho R} \right) \boldsymbol{(v_{g}-v)}  & R\ll\zeta\\
    \left(\dfrac{\rho_{g}B}{\rho R} \right) \boldsymbol{(v_{g}-v)\vert v_{g}-v\vert} &  R\gg\zeta
  \end{cases}
\end{equation} 
where $\rho_{g}$ is the density of the gas, $\zeta$ its mean free path length, $\boldsymbol{v_{g}}$ its velocity, $v_{s}$ the local sound speed and B is given by the equation 
\begin{equation}  
B=\begin{cases}
    9\left[ \dfrac{6R}{\zeta v_{R}} \boldsymbol{\vert v_{g}-v\vert} \right]^{-1}  &  Re\leq 1\\
    9\left[ \dfrac{6R}{\zeta v_{R}} \boldsymbol{\vert v_{g}-v\vert} \right]^{-0.6} &  1\leq Re\leq 800 \\
    0.165 &  Re\geq 800
  \end{cases}
\end{equation} 
for differing values of the Reynolds number, calculated from the equation 
\begin{equation}  
Re=\dfrac{6R}{\zeta v_{s}}\boldsymbol{\vert v_{g}-v\vert}
\end{equation}  
For the mean free path, we use the approximate relation from \cite{Veras2015}
\begin{equation}  
\rho_{g}\zeta\sim10^{-8} \text{kg\,m}^{-2}
\end{equation}  

The stellar wind in our simulations was assumed to travel radially and with the escape velocity of the system. The mass loss during the CE will occur preferentially in the disc plane, and we therefore confine all stellar wind to a disc of scale height $H=0.15r$. However, this preferential direction also suggests the particles will not be expelled perfectly radially. We therefore assume only 15\% of the expelled mass travels radially outwards and interacts with the particles, which should ensure our drag force is a conservative estimate.

%%%%%%%%%%%%%%%%%%%%%%%%%%%%%%%%%%%%%%%%%%%%%%%%%%

% Don't change these lines
\bsp	% typesetting comment
\label{lastpage}
\end{document}